# Mobile Zigbee Sensor Networks

Er.Anantdeep, Er. Sandeep kaur and Er.Balpreet Kaur

**Abstract**— OPNET Modeler accelerates network R&D and improves product quality through high-fidelity modeling and scalable simulation. It provides a virtual environment for designing protocols and devices, and for testing and demonstrating designs in realistic scenarios prior to production. OPNET Modeler supports 802.15.4 standard and has been used to make a model of PAN. Iterations have been performed by changing the Power of the transmitter and the throughput will has been analyzed to arrive at optimal values.An energy-efficient wireless home network based on IEEE 802.15.4, a novel architecture has been proposed. In this architecture, all nodes are classified into stationary nodes and mobile nodes according to the functionality of each node. Mobile nodes are usually battery-powered, and therefore need low-power operation. In order to improve power consumption of mobile nodes, effective handover sequence based on MAC broadcast and transmission power control based on LQ (link quality) are employed. Experimental results demonstrate that by using the proposed architecture, communication time and power consumption of mobile nodes can be reduced by 1.2 seconds and 42.8%, respectively.

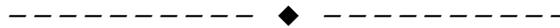

## 1 INTRODUCTION

ZigBee protocols are intended for use in embedded applications requiring low data rates and low power consumption. ZigBee's current focus is to define a general-purpose, inexpensive, self-organizing mesh network that can be used for industrial control, embedded sensing, medical data collection, smoke and intruder warning, building automation, home automation, etc. The resulting network will use very small amounts of power -- individual devices must have a battery life of at least two years to pass ZigBee certification. Typical application areas include : Home Entertainment and Control — Smart lighting, advanced temperature control, safety and security, movies and music, Home Awareness — Water sensors, power sensors, smoke and fire detectors, smart appliances and access sensors, Mobile Services — m-payment, m-monitoring and control, m-security and access control, m-healthcare and tele-assist, Commercial Building — Energy monitoring, HVAC, lighting, access control, Industrial Plant — Process control, asset management, environmental management, energy management, industrial device control.

## 2 TYPES OF ZIGBEE DEVICES

There are three different types of ZigBee devices: ZigBee coordinator(ZC): The most capable device, the coordinator forms the root of the network tree and might bridge to other networks. There is exactly one ZigBee coordinator in each network since it is the device that started the network originally. It is able to store information about the network, including acting as the Trust Centre & repository for security keys.

ZigBee Router (ZR): As well as running an application function a router can act as an intermediate router, passing data from other devices.

ZigBee End Device (ZED): Contains just enough functionality to talk to the parent node (either the coordinator or a router); it cannot relay data from other devices. This relationship allows the node to be asleep a significant amount of the time thereby giving long battery life. A ZED requires the least amount of memory, and therefore can be less expensive to manufacture than a ZR or ZC.

The protocols build on recent algorithmic research (Ad-hoc On-demand Distance Vector, neuRFon) to automatically construct a low-speed ad-hoc network of nodes. In most large network instances, the network will be a cluster of clusters. It can also form a mesh or a single cluster. The current profiles derived from the ZigBee protocols support beacon and non-beacon enabled networks. In non-beacon-enabled networks (those whose beacon order is 15), an unslotted CSMA/CA channel access mechanism is used. In this type of network, ZigBee Routers typically have their receivers continuously active, requiring a more robust power supply. However, this allows for heterogeneous networks in which some devices receive continuously, while others only transmit when an external stimulus is detected. The typical example of a heterogeneous network is a wireless light switch: the ZigBee node at the lamp may receive constantly, since it is connected to the mains supply, while a battery-powered light switch would remain asleep until the switch is thrown. The switch then wakes up, sends a command to the lamp, receives an acknowledgment, and returns to sleep. In such a network the lamp node will be at least a ZigBee Router, if not the ZigBee Coordinator; the switch node is typically a ZigBee End Device. In beacon-enabled networks, the special network nodes called ZigBee Routers transmit periodic beacons to confirm their presence to other network nodes. Nodes may sleep between beacons, thus lowering their duty cycle and extending their battery life. Beacon intervals may range from 15.36 milliseconds to 15.36 ms * 214 = 251.65824 seconds at 250 kbit/s, from 24 milliseconds to 24 ms * 214 = 393.216 seconds at 40 kbit/s and from 48 milliseconds to 48 ms * 214 = 786.432 seconds at 20 kbit/s. However, low duty cycle operation with long beacon intervals requires precise timing, which can conflict with the need for low product cost. In general, the ZigBee protocols minimize the time the radio is on so as to reduce power use. In beaconing networks, nodes only need to be active while a beacon is being transmitted. In non-beacon-enabled networks, power consumption is decidedly asymmetrical: some devices are always active, while others spend most of their time sleeping. ZigBee devices are required to conform to the IEEE 802.15.4-2003 Low-Rate Wireless Personal Area Network (WPAN) standard. The standard specifies the lower protocol layers—the physical layer (PHY), and the medium access control (MAC) portion of the data link layer (DLL). This standard specifies operation in the unlicensed 2.4 GHz, 915 MHz and 868 MHz ISM bands. In the 2.4 GHz band there are 16 ZigBee channels, with each channel requiring 5 MHz of bandwidth. The center frequency for each channel can be calculated as, FC = (2350 + (5 * ch)) MHz, where ch = 11, 12, ..., 26. The radios use direct-sequence spread spectrum coding, which is managed by the digital stream into the modulator. BPSK is used in the 868 and 915 MHz bands, and orthogonal QPSK that transmits two bits per symbol is used in the 2.4 GHz band. The raw, over-the-air data rate is 250 kbit/s per channel in the 2.4 GHz band, 40 kbit/s per channel in the 915 MHz band, and 20 kbit/s in the 868 MHz band. Transmission range is between 10 and 75 meters (33 and 246 feet), although it is heavily dependent on the particular environment. The maximum output power of the radios is generally 0 dBm (1 mW). The basic channel access mode is carrier sense, multiple access/collision avoidance, (CSMA/CA). That is, the nodes talk in the same way that people converse; they briefly check to see that no one is talking before they start. There are three notable exceptions to the use of CSMA. Beacons are sent on a fixed timing schedule, and do not use CSMA. Message acknowledgments also do not use CSMA. Finally, devices in Beacon Oriented networks that have low latency real-time requirements may also use Guaranteed Time Slots (GTS), which by definition do not use CSMA.The software is designed to be easy to develop on small, cheap microprocessors. The radio design used by ZigBee has been carefully optimized for low cost in large scale production. It has few analog stages and uses digital circuits wherever possible. Even though the radios themselves are cheap, the ZigBee Qualification Process involves a full validation of the requirements of the physical layer. This amount of concern about the Physical Layer has multiple benefits, since all radios derived from that semiconductor mask set would enjoy the same RF characteristics. On the other hand, an uncertified physical layer that malfunctions could cripple the battery lifespan of other devices on a ZigBee network. Where other protocols can mask poor sensitivity or other esoteric problems in a fade compensation response, ZigBee radios have very tight engineering constraints: they are both power and bandwidth constrained. Thus, radios are tested to the ISO 17025 standard with guidance given by Clause 6 of the 802.15.4-2006 Standard. Most vendors plan to integrate the radio and microcontroller onto a single chip. Bluetooth and Wi-Fi were both rejected due to their power consumption, cost, complexity and size.

## 3. COMPARISON

The table 1 shows the Zigbee protocol stack size to be 32 KB. This is less than one-third of the stack size required by Wi-Fi and Bluetooth.Around 1998, a Zigbee-style network was most popular. This comprised of a self-organising ad-hoc network of sensors. As a result, the IEEE 802.15.4 standard was drawn up and completed in May 2003. This standard documents a set of specifications for the MAC (Medium Access Control) and PHY (Physical Layer) for LR-WPANs ( Low Rate Wireless Personal Area Networks).

More specifically, the IEEE 802.15.4 standard specifies the PHY and MAC layers at the 868 MHz, 915 MHz and 2.4 GHz ISM Bands. This enables global or regional deployment. The 2.4 GHz band has 16 channels and is supported globally. The 915 MHz, with 10 channels, is supported in the Americas. The 868 MHz with 1 channel is supported in Europe. The data rates for each are 250, 40 and 20 kbps respectively. The 2.4 GHz offers a range of up to 30 meters and a transmission rate of 250 Kbps. The Zigbee standard is founded on the IEEE 802.15.4 standard. It enables the interoperability of wireless devices in a network. The Zigbee alliance, an industry working group, is working to standardize the application software that sits on top of the IEEE 802.15.4 layers. The purpose is to give rise to a broad range of interoperable consumer devices. The Zigbee 1.0 specifications were ratified on 14 December 2004. The Zigbee architecture is illustrated below.

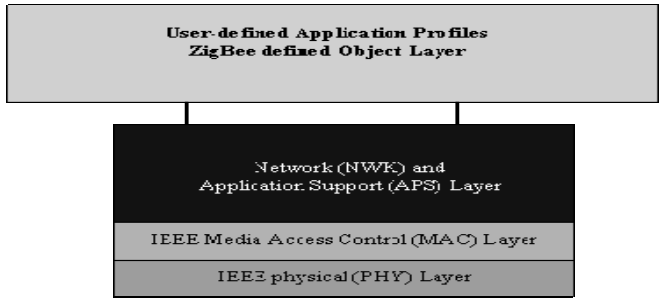

Figure 1 IEEE 802.15.4 / Zigbee Stack Architecture

The architecture is divided into layers, as can be seen in the diagram. The network layer allows for the growth of a network. It can support a large number of nodes. This layer includes Zigbee device object (ZDO), the user-defined application and the application support (APS) sublayer

The APS is involved in the communication between devices, the initialisation of a devicein a network, the routing of messages and discovery. Discovery is the identifying of other devices in the operating space of any device. It is also concerned with binding. Binding is the matching of different but compatible devices together, such as a lamp and a switch.

The responsibility of the ZDO is in dealing with bind requests, ensuring security between devices and determining the device type. The device type refers to the capabilities of the device and its complexity. The user-defined application is the Zigbee compliant end device.

The MAC layer is concerned with single-hop communication between neighbouring devices. It synchronises the network and provides reliability between two devices. The physical layer (PHY) defines radio characteristics.

Table 1: Comparision of different wireless technologies

| Standard | Bandwidth | Power Consumption | Protocol Stack Size | Stronghold |
|---|---|---|---|---|
| Wi-Fi | Up to 54Mbps | 444+mA TX, standby 20mA | 100+KB | High data rate |
| Bluetooth | 1Mbps | 40mA TX, standby 0.2mA | ~100+KB | Interoperability, cable replacement |
| Zigbee | 250kbps | 30mA TX, standby 3μA | 32KB | Long battery life, low cost |

## 4 RESULTS AND DISCUSSIONS

The output of the simulation testbed is as shown below. In the first trace the power of the transmitters has been fixed at 0 dBm. It can be seen that there is an interval in which the mobile node remains out of communication ie from 2 to 4 m and from 11 to 13 m.

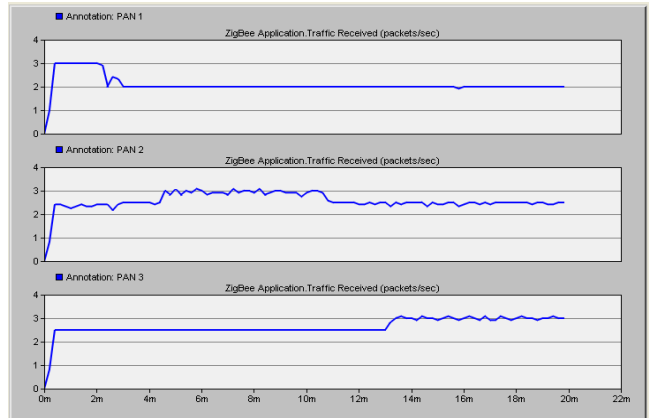

Figure 2(a): First Trace

In the second trace the power of the transmitters has been fixed at 0 dBm. It can be seen that there is an interval in which the mobile node remains out of communication has reduced but still this is not the optimum power level.

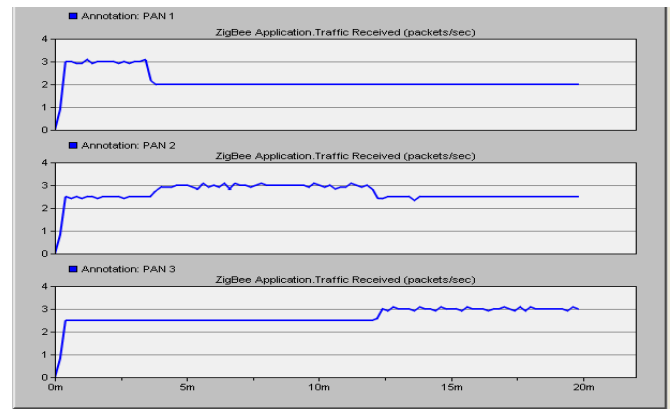

Figure 2(b) Second Trace

In the third trace the power of the transmitters has been fixed at 3 dBm. It can be seen that there is an interval in which the mobile node remains out of communication has reduced but still this is not the optimum power level.

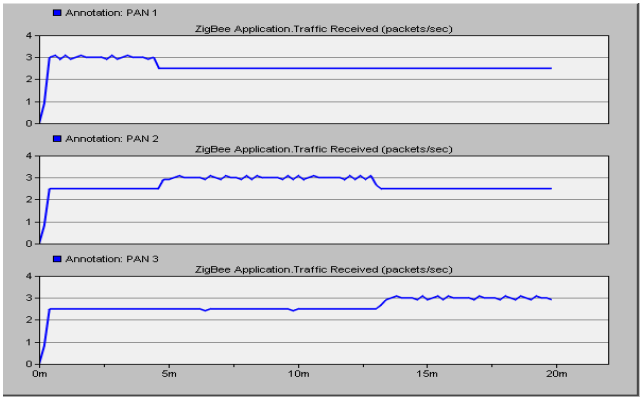

Figure 2(c) Third Trace

In the fourth trace the power of the transmitters has been fixed at 4 dBm. It can be seen that there is an interval in which the mobile node remains in communication throughout. So this is the optimal power level for Tx.

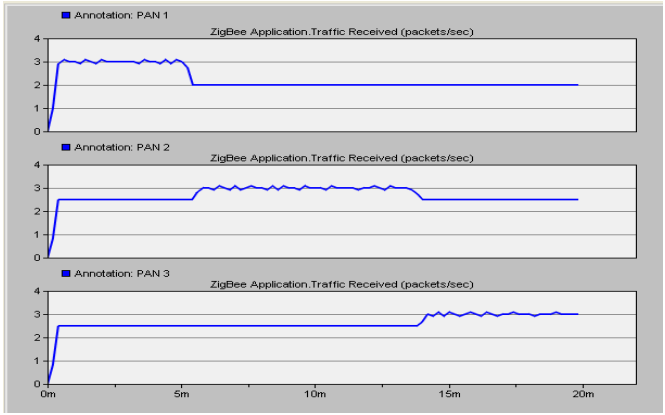

Figure 2(d) Fourth Trace

In the fifth trace the power of the transmitters has been fixed at 5 dBm. It can be seen that there is an interval in which the mobile node remains in communication with multiple nodes and over a longer sidtance so now the power level is more than the desirable and is not the optimum power level .

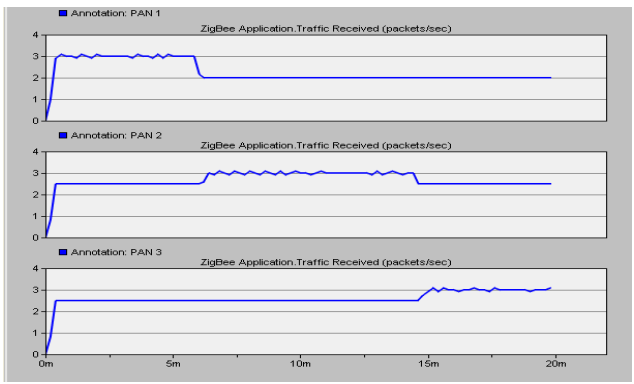

Figure 2(e) Fifth Trace

In the sixth trace the power of the transmitters has been fixed at 6 dBm. It can be seen that there is an interval in which the mobile node remains in communication with multiple nodes and over a longer distance so now the power level is more than the desirable and is not the optimum power level .

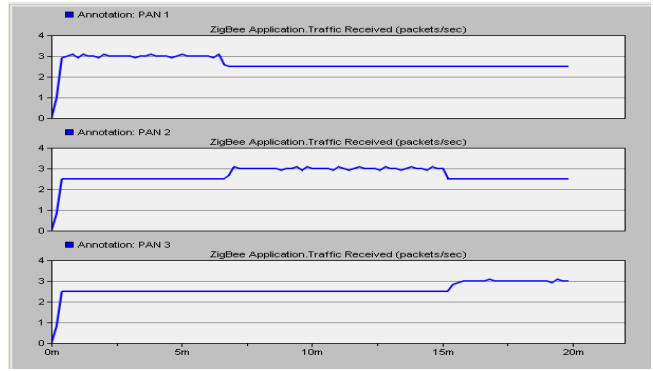

Figure 2(f) Sixth Trace

## 5 Conclusions

Thus it can be seen that a testbed for optimizing the power output of zigbee network has been successfully implemented.  In the first trace the power of the transmitters has been fixed at 0 dBm. It can be seen that there is an interval in which the mobile node remains out of communication ie from 2 to 4 m and from 11 to 13 m. In the second trace the power of the transmitters has been fixed at 0 dBm. It can be seen that there is an interval in which the mobile node remains out of communication has reduced but still this is not the optimum power level In the third trace the power of the transmitters has been fixed at 3 dBm. It can be seen that there is an interval in which the mobile node remains out of communication has reduced but still this is not the optimum power level In the fourth trace the power of the transmitters has been fixed at 4 dBm. It can be seen that there is an interval in which the mobile node remains out of communication has reduced but still this is not the optimum power level In the fifth trace the power of the transmitters has been fixed at 5 dBm. It can be seen that there is an interval in which the mobile node remains out of communication has reduced but still this is not the optimum power level .In the sixth trace the power of the transmitters has been fixed at 6dBm. It can be seen that there is an interval in which the mobile node remains out of communication has reduced but still this is not the optimum power level level.

## 6 FUTURE WORK

Testbed can be modified to optimize power levels for other networks such as Blue-tooth

## ACKNOWLEDGMENT

We are very thankful to Dr. Amardeep, Reader, University College of Engineering (UCOE), Punjabi University, Patiala for giving us such a valuable support in doing thesis work.

## REFERENCES

[1] Betta G.; Capriglione D.; Ferrigno L.; Miele G.; "Experimental Investigation of the Electromagnetic Interference of ZigBee Transmitters on Measurement Instruments" Instrumentation and Measurement, IEEE Transactions on Volume 57, Issue 10, Oct. 2008 Page(s):2118 - 2127

[2] Cheolhee Park; Rappaport ; T.S.; "Short-Range Wireless Communications for Next-Generation Networks: UWB" 60 GHz Millimeter-Wave WPAN And ZigBee Wireless Communications, IEEE [see also IEEE Personal Communications]Volume 14, Issue 4, August 2007 Page(s):70 - 78

[3] Chonggang Wang;Sohraby K; Jana, R.; Lusheng Ji; Daneshmand M.; "Voice communications over zigbee networks" Communications Magazine, IEEE,Volume 46, Issue 1, January 2008 Page(s):121 - 127

[4] Chua-Chin Wang; Chi-Chun Huang; Jian-Ming Huang; Chih-Yi Chang; Chih-Peng Li; "ZigBee 868/915-MHz Modulator/Demodulator for Wireless Personal Area Network" Very Large Scale Integration (VLSI) Systems, IEEE Transactions on Volume 16, Issue 7, July 2008 Page(s):936 - 939

[5] Guthrie B.; Hughes J.; Sayers T.; Spencer A.; "A CMOS gyrator low-IF filter for a dual-mode Bluetooth/ZigBee transceiver Solid-State Circuits" IEEE Journal ,Volume 40, Issue 9, Sept. 2005 Page(s):1872 - 1879

[6] Gershenfeld N.; Cohen D.; "Internet 0: Interdevice Internetworking - End-to-End Modulation for Embedded Networks " Circuits and Devices Magazine, IEEE Volume 22, Issue 5, Sept.-Oct. 2006 Page(s):48 - 55

[7] Hsin-Mu Tsai; Tonguz O.K.; Saraydar C.; Talty T.; Ames M.; Macdonald A.; "Zigbee-based intra-car wireless sensor networks: a case study" Wireless Communications, IEEE [see also IEEE Personal Communications Volume 14, Issue 6, December 2007 Page(s):67 - 77

[8] Ishibashi, K.; Ochiai H.; Kohno; "Embedded forward error control technique (EFECT) for low-rate but low latency communications" R.Wireless Communications, IEEE Transactions on Computing,Volume 7, Issue 5, Part 1, May 2008, Page(s):1456 - 1460

[9] Jianliang Zheng; Lee, M.J.; Anshel, M.; "Toward Secure Low Rate Wireless Personal Area Networks" IEEE Transactions on Computing,Volume 5, Issue 10, Oct. 2006 ,Page(s):1361 - 1373

[10] Joon Heo; Choong Seon Hong; Seok Bong Kang; Sang Soo Jeon; "Design and Implementation of Control Mechanism for Standby Power Reduction,Consumer Electronics", IEEE Transactions on Volume 54, Issue 1, February 2008, Page(s):179 - 185

[11] Jae Yeol Ha; Hong Seong Park; Sunghyun Choi; Wook Hyun Kwon;EHRP: Enhanced hierarchical routing protocol for zigbee mesh networks Communications Letters" IEEE ,Volume 11, Issue 12, December 2007 Page(s):1028 - 1030

[12] K.E; Chi-Chih Chen; Volakis; "A Novel Radiator for a 2.4 GHz Wireless Unit to Monitor Rail Stress and Strain Browne",J.LAntennas and Propagation, IEEE Transactions on Computing,Volume 56, Issue 3, March 2008 Page(s):887 – 892.

**Anantdeep**, presently working as Lecturer, Department of CSE/IT, BBSB Engineering College, Fatehgarh Sahib. She is M-Tech in ICT from punjabi university patiala. Her major research interests include mobile zigbee networks. Also Anantdeep is having 4 publications in various National and International Conferences and/or journals.

**Sandeep Kaur**, presently working as Lecturer in the Department CSE/IT of BBSB Engineering College, Fatehgarh Sahib, Punjab (INDIA). She is master of engineering in computer science & engineering from Thapar University Patiala in 2009. Her major research interests parallel computing and performance of optical multistage interconnection networks. Also Sandeep kaur is having 8 publications in various National and International Conferences.

**Balpreet Kaur**, presently working as Lecturer, Department of CSE/IT, BBSB Engineering College, Fatehgarh Sahib,PUNJAB(INDIA). She is persuing M-Tech in BBSBEC ,Fatehgarh sahib. Her major research interests include mobile zigbee networks. Also Balpreet kaur is having 5 publications in various National and International Conferences.